\documentclass[reprint,amsmath,amssymb,aip,jcp,author-numerical]{revtex4-1}
\usepackage{graphics}
\usepackage{graphicx}
\usepackage{dcolumn}
\usepackage{bm}
\usepackage{natbib}
\usepackage{color}
\usepackage{hyperref}

\begin{document}

\title{Fluctuations in reactive networks subject to extrinsic noise studied in the framework of the Chemical Langevin Equation.}

\author{H. Berthoumieux}

\affiliation{CNRS, UMR 7600, LPTMC, F-75005, Paris, France}
\affiliation{Sorbonne Universit\'es, UPMC Univ Paris 06, UMR 7600, LPTMC, F-75005, Paris, France}

\begin{abstract} 
Theoretical and experimental studies have shown that the fluctuations of {\it in vivo} systems break the fluctuation-dissipation theorem. One can thus ask what information is contained in the correlation functions of protein concentrations and how they relate to the response of the reactive network to a perturbation. Answers to these questions are of prime importance to extract meaningful parameters from the {\it in vivo} fluorescence correlation spectroscopy data. In this paper we study the fluctuations of the concentration of a reactive species involved in a cyclic network that is in a non-equilibrium steady state perturbed by a noisy force, taking into account both the breaking of detailed balance and extrinsic noises.  Using a generic model for the network and the extrinsic noise, we derive a Chemical Langevin Equation that describes the dynamics of the system, we determine the expressions of the correlation functions of the concentrations, estimate the deviation of the fluctuation-dissipation theorem and the range of parameters in which an effective temperature can be defined.

\end{abstract}
\maketitle

\section{Introduction}
For a chemist, the living world is a system capable to convert an energy source into self-organization. In a cell, biochemical pathways are organized to transmit a signal from outside of the cell to the nucleus, to polarize the cell, etc... \cite{siowling2010}. The networks are out-of-equilibrium and energy sources (sources of reactants, sinks of products, temperature, light, ... )  drive the reactive cycles in a particular direction. The kinase signaling cascade that accelerates and amplifies a signal from the cell membrane to the nucleus functioning as a domino-like relay is an example of these interconnected out-of-equilibrium cycles \cite{kholodenko2006}.
 
To highlight the key ingredients governing the self-organization in cells, different approaches have been undertaken. Biomimetic approaches consist in coupling patterns of energy sources with reaction-diffusion systems in order to reproduce organization states of biologic interest \cite{lesaux2014}. In parallel to the design of artificial systems, quantitative description of {\it in vivo} cellular protein networks have been developped \cite{machan2014,wachsmuth2015}. Fluorescence correlation spectroscopy (FCS) is a single molecule method that gives acces to the abundance, the diffusive and the reactive properties of biochemical species by analyzing local concentration fluctuations  \cite{mütze2011}. This technique was designed more than 40 years ago \cite{madge1972} and was generalized during the last ten years for the study of living systems. The analysis of FCS data involves two steps,  the determination of the chemical mechanism (reaction, diffusion, convection, ...) which is {\it a priori} unknown and the estimation of the parameters of the model \cite{he2012,guo2013}. However, the theoretical framework used in the fitting procedure remains the one developed in the seminal paper based on equilibrium thermodynamics fluctuations\cite{elson1974}.   
 
Important effort has also been made to describe out-of-equilibrium systems. The field of stochastic thermodynamics has allowed a better comprehension by introducing the concepts of internal energy, heat, work \cite{sekimoto2010} and entropy \cite{seifert2005} at the level of a trajectory, {\it i. e.} as fluctuating quantities. Important properties of their probability distributions were established (see \cite{seifert2012} for a review). For systems in a non-equilibrium steady state (NESS), modified fluctuation-dissipation relations were derived \cite{harada2005,speck2006,chetrite2008}. These were applied to small enzymatic cycles maintained out of equilibrium by sources and sinks of substrates and products \cite{schmiedl2007,verley2011}. Such relations provide a frame to extract the information contained in correlation functions of species involved in out-of-equilibrium networks.

A model of a chemical cycle maintained in a NESS by a constant driving force does not include the fluctuations generated by the environment. However, extrinsic fluctuations are omnipresent in a living cell\cite{elowitz2002,paulsson2004}. An example often discussed in the literature is the translation process: the number of protein copies is a stochastic variable whose probability distribution depends on the distribution of the number of messenger RNA and on the intrinsic noise originating in the stochastic nature of the translation process. The first one, referred to as extrinsic noise, was shown to represent the main source of concentration variance in many biological pathways \cite{taniguchi2010,jones2014}.
The effect of the input noise in the signaling function of genetic circuits has been extensively studied \cite{tanasenicola2006,hansen2015}.   

In this paper, we consider a three-state chemical cycle in a NESS perturbed by a noisy force with vanishing value. In this case, the extrinsic fluctuations do nut blur an input signal received by the network but perturb an energy source that maintains the system out of equilibrium. How does this extrinsic noise influence the variance and the correlation of the chemical species concentrations and what are the consequences on the modified fluctuation-dissipation relations? Here, we describe the dynamics of the system is described in the frame of Chemical Langevin Equations obtained by a coarse-graining on a small macroscopic time $\tau$ \cite{gillespie2000}. We show that a fast relaxing extrinsic noise mimics a thermal noise and modifies the diffusion term of the Langevin equation whereas a slowly relaxing noise generates an extra multiplicative force that fluctuates at the extrinsic noise timescales.

The paper is organized as follows. In the first part, we present the model used for the chemical cycle and the driving force. In the second part, following Gillespie's paper \cite{gillespie2000}, the Chemical Langevin Equations are established and a particular attention is paid to the noise term which depends on the driving force fluctuations timescale. In the third part, the modified fluctuation-dissipation theorem is derived in the case of a constant driving force and the effects of extrinsic fluctuations on this relation are discussed.
 The last parts are devoted to the discussion and the conclusion. 

\section{Model}
\begin{figure}
\includegraphics[scale=1]{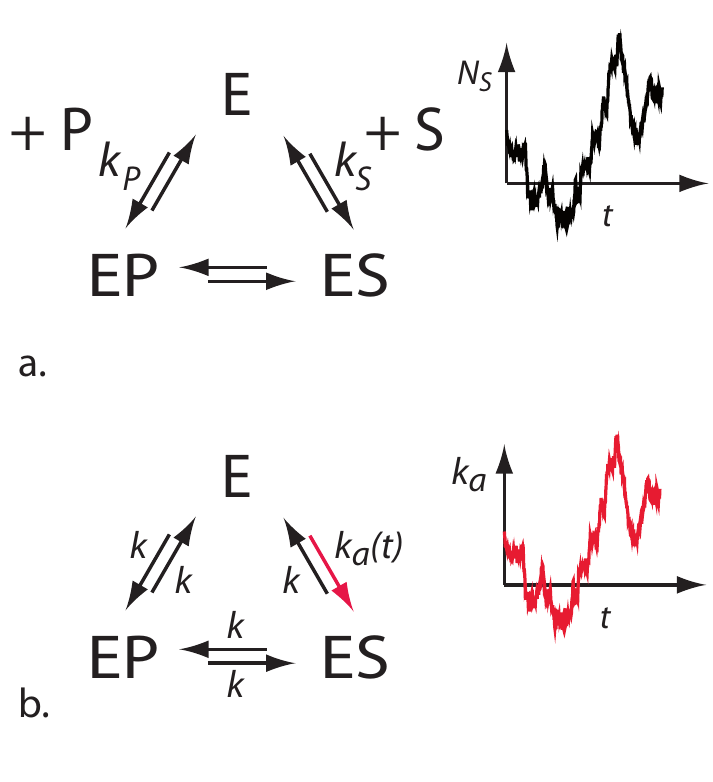} 
\caption{Cyclic three-state Michaelis-Menten mechanism maintained out-of-equilibrium. The enzyme E catalyses the transformation of the substrate S in the product P. The concentrations of both S and P are controlled by mechanisms external to the catalytic cycle and $s(t)$ follows the dynamics of an Onrstein-Uhlenbeck process governed by Eq. (\ref{O-U}). This chemical system can be modeled by a set of pseudo-isomerizations, one of which ( E $\rightarrow$ ES) being associated with a time-dependent rate constant $k_a(t)$ given in Eq. (\ref{ka}).}
\end{figure}
We consider the transformation of a substrate S in a product P catalyzed by an enzyme E. This reaction is modeled by a three-state Michaelis-Menten mechanism represented in Fig. 1. 
The cycle is composed of four isomeriations, ES $\rightarrow$ E, ES $\rightarrow$ EP, EP $\rightarrow$ ES, EP $\rightarrow$ E, that for sake of simplicity are characterized by the same rate constant $k$, and of two second order reactions E+S $\rightarrow$ ES, E+P $\rightarrow$ EP associated with the rates $k_S$ and $k_P$ respectively.
The substrate and the product concentrations, $S(t)$ and $P(t)$, are supposed to be controlled by other processes in the cell and act as external driving forces that maintain the cycle in a NESS. 
The expression of the stochastic thermodynamics quantities, such as heat, entropy production, ..., and their related properties  have been derived for this system \cite{schmiedl2007,ge2012}. Moreover, this cycle is a toy model used to develop  methods of mechanism identification \cite{berthoumieux2007,berthoumieux2009,lemarchand2012} and of discrimination between equilibrium and non-equilibrium steady-states for {\it in vivo} systems \cite{qian2004,berthoumieux20092,bianca2014}. 
   
We consider the case in which $P(t)$ is constant and fixed to $P=k/k_P$, and $S(t)$ is fluctuating following an Ornstein-Uhlenbeck process. We define the rescaled concentration $s(t)=k_S S(t)/k$ whose dynamics is governed by 
\begin{equation}
\label{O-U}
\frac{d s(t)}{dt}=-\frac{s(t)-s_m}{\tau_s} +\sqrt{\frac{2}{\tau_s}}\sigma_s\Gamma(t),
\end{equation}
with a relaxation time $\tau_s$ and a diffusion coefficient $\sqrt{\frac{2}{\tau_s}}\sigma_s$.
$\Gamma(t)$ is a white noise of unit variance. In the stationary state, the substrate concentration $s(t)$ obeys a normal distribution
\begin{equation}
\label{PS}
\mathcal{P}(s)=\frac{1}{\sqrt{2\pi}\sigma_s}e^{-\frac{(s-s_m )^2}{2\sigma_s^2}},
\end{equation}
of mean value $s_m$ and of variance $\sigma_s$.  Note that $s(t)$ has to be positive which imposes the condition $\sigma_s/s_m \ll 1$.
The concentration $s(t)$ is equal to
\begin{eqnarray}
s(t)&=&s_m+\delta s(t),\\  {\rm with} \quad \delta s(t)&=&\int_{-\infty}^tdt'e^{t'/\tau_s}\sqrt{\frac{2}{\tau_s}}\sigma_s\Gamma(t')e^{-t/\tau_s}.
\end{eqnarray}
In the limit of a vanishing variance $\sigma_s$, the concentration
$s(t)$ is constant and equal to $s_m$.   

The enzymatic cycle can be represented by the scheme given in Fig. 1b, a set of six isomerizations driven in a non-equilibrium steady state by a fluctuating chemical rate,  
\begin{equation}
\label{ka}
k_a(t)=ks(t)=k_a+\delta k_a(t),
\end{equation}
of mean value $k_a=k s_m$ and fluctuating part $\delta k_a(t)=k\delta s(t)$, with $\langle \delta k_a(t) \rangle=0$. 

The chemical system is in an equilibrium state in the absence of external driving and extrinsic fluctuations, conditions that are expressed by the vanishing values of two dimensionless parameters $(\lambda=0, \Delta=0)$, where  
\begin{equation}
\lambda=\log (s_m), \quad \Delta=\frac{\sigma_s}{s_m}.
\end{equation}
$\lambda$ represents the chemical potential difference between S, P in $k_BT$ units and $\Delta$ is the rescaled width of the probability distribution of $s$.   

A system composed of $N$ molecules maintained in a NESS by a non-noisy driving force ($\lambda\neq 0$,$\Delta=0$) is characterized by a mean number of particles in configurations E and ES such that \cite{berthoumieux20092}
\begin{eqnarray}
\label{station}
\langle N_E \rangle &=& P_E N, \quad \langle N_{ES} \rangle= P_{ES} N \\
\label{proba}
{\rm with} \quad P_E &=&\frac{1}{2+s_m}, \quad P_{ES}=\frac{2s_m+1}{3(2+s_m)}.
\end{eqnarray} 
A positive (negative), value of $\lambda$ corresponds to a reaction flux in the direction S $\rightarrow $ P (P$\rightarrow$S). 
 The parameter space $(\lambda\neq0,\Delta>0)$ models a chemical system in a NESS perturbed by a noisy source of energy. This will generates non-equilibrium intrinsic noises ($\lambda \neq 0$) and extrinsic noises ($\Delta >0 $). We aim to characterize the fluctuations of $N_i(t)$, $(i=E,ES,EP)$, for such systems. To do so, we derive in the next section the Chemical Langevin Equations describing the dynamics of this system.

To check our analytic results, we simulate numerically the evolution of the system represented in Fig. 1. The enzymatic cycle dynamics is described by a master equation associated with time-dependent rates which is solved using the Gillespie algorithm\cite{gillespie1976}. For each time $t$ at which one of the six possible chemical transformation occurs, the concentration of the substrate $s(t)$ is updated using the equation for an Ornstein-Uhlenbeck process \cite{gillespie1996}.

\section{Chemical Langevin Equation}
In this part, we derive the Chemical Langevin Equations that 
govern the dynamics of the network described in Fig. 1b and we calculate the correlation functions associated with these equations.

We assume that the total number of enzyme $N$ is constant. At time $t$, $N_E(t)$ molecules are in the conformation E, $N_{ES}(t)$ in the conformation ES. The number $N_{EP}(t)$ of enzymes in state EP is deduced from the matter conservation $N_{EP}(t)=N-N_{E}(t)-N_{ES}(t)$.
The chemical system is thus in a state $e(t)=\left( \begin{array}{c} N_E(t) \\ N_{ES}(t)  \end{array} \right)$. 
Let $n_r^j(e(t), \tau)$,  for any $\tau>0$, be the number of reactions $j$ ($j=$E$\rightarrow$ ES, ES$\rightarrow$ E ..., EP$\rightarrow$ E) that occur during the time interval $[t,t+\tau]$. The chemical state at $t+\tau$, $e(t+\tau)$  can be expressed as
\begin{eqnarray}
\label{propeq}
e_i(t+\tau)&=&e_i(t)+n_r^j(e(t),\tau)\nu_{ij},\quad i=(1,2),\nonumber\\
 j&=&E\rightarrow ES, ..., EP\rightarrow E                                                  
     \end{eqnarray} 
where $\nu_{ij}$ is the stoichiometric coefficient associated with  species $i$ in the reaction channel $j$. For a channel associated with a constant rate, {\it i. e.} $j \neq {\rm E}\rightarrow {\rm ES}$, the number of reactions $n_r^j(e(t),\tau)$ is a stochastic variable that follows a Gaussian probability law for unrestrictive conditions, as demonstrated by Gillespie in reference \cite{gillespie2000}.

The first condition is that the time interval $\tau$ should be small enough so that the change in the state $e$ during $[t, t+\tau]$ does not change $n^r_j$ significantly. This means that 
\begin{equation}
\label{cond1}
N_i(t') \approx N_i(t) \quad \forall t' \in [t,t+\tau], \quad i=(E, ES, EP).    
\end{equation}
In other words, the number of reactions that occur during $\tau$ should be much smaller than the number of molecules in each state. Such a condition can be fulfilled by choosing a small time interval. In this approximation, the number of chemical transformations will be a statistically independent random variable that follows a Poisson law $P_j(\Lambda^j)$ of rate $\Lambda^j$ which expression depends of the reaction channel, for example $\Lambda^{ES \rightarrow EP}=kN_{ES}(t)\tau$. The mean value $\langle n_r^{ES \rightarrow EP} \rangle$ of $n^{ES\rightarrow EP}_r$ is equal to $kN_{ES}(t)\tau$, its variance $\sigma_r^{ES\rightarrow EP}$ is equal to $\sqrt{kN_{ES}(t)\tau}$.

The second condition is to choose $\tau$ large enough so that the number of occurrence of each reaction is much larger than one,
\begin{equation}
\label{cond2}
\langle n_r^j (\tau) \rangle \gg 1 \quad \forall j.
\end{equation} 
In this case, the Poisson variable $n_r^j$ can be approximated by a variablefollowing a normal probability law $\mathcal{N}_j(\langle n_r^j \rangle ,\sigma_r^j)$ of mean value $\langle n^r_j \rangle$ and variance $\sigma_r^j$ \cite{gillespie2000} . The random variable $n_r^j$ can be written as follows,
\begin{equation}
n_r^j=\langle n_r^j\rangle +\sigma_r^j \delta^j,
\end{equation}
where $\delta^j$ is a normal random variable of vanishing mean value and unit variance.
A large number of particles in each state is in most cases a sufficient condition for the existence of a time $\tau$ fulfilling the conditions given by Eqs. (\ref{cond1},\ref{cond2}). $\tau$ is a mesoscopic time interval large enough so that each reaction channel is visited several times but small enough so that the state of system does not change significantly. The following expressions are valid for a time coarse-grained on $\tau$.

Once this time is estimated for the channels associated with constant rates, we consider the chemical transformation E$\rightarrow$ES associated with a fluctuating rate. Two situations are envisaged. First, $\tau_s$ the characteristic time of the Ornstein-Uhlenbeck process defined in Eq. (\ref{O-U}) is much smaller than $\tau$ ($\tau\gg\tau_s$). In this case, $s(t)$ and $k_a(t)$ reach a stationary distribution on $\tau$. In the second case, $\tau_s$ is much larger than $\tau$, $s(t)$ and $k_a(t)$ are constant on $\tau$ but vary on longer timescales ($\tau\ll\tau_s$). 
In both cases, the number of reactions $n_r^{E \rightarrow ES}(\tau)$ can be approximated by a normal variable  which mean and variance is given in Table I (See appendix A for the derivation).

\begin{table*}

\center
\renewcommand{\arraystretch}{1.6}
\setlength{\tabcolsep}{1cm}
\begin{tabular}{cccc}

\hline
 & &$\langle n_r \rangle= d_r N_E(t)\tau$ & $\sigma_r^2=v_r^2\tau$ \\
& &with $d_r$ & with $v_{r}^2$ \\ 
\hline
\hline
Constant rate &  & $k$ & $kN_E(t)$ \\  
 ES $\rightarrow$E & &  & \\
\hline
\hline
Fluctuating rate & $\Delta=0$ & $ks_m $  & $kN_E(t)s_m$  \\
\cline{2-4}
E $\rightarrow$ ES  & $\tau_S \ll \tau$ & $ks_m $ & $(kN_E(t)s_m + 2k^2N_E^2(t)\sigma_s^2\tau_s)$   \\
\cline{2-4}
& $\tau_S \gg \tau$ &$k s(t) $  & $ kN_E(t)s(t) $   \\
\hline 
\hline

\end{tabular}
\caption{Mean value and variance of the reaction number $n_r$ during the mesoscopic time $\tau$ for constant and fluctuating rates. 
For fluctuating rates, the two cases of a fast and a slow Ornstein-Uhlenbeck process are considered. The expressions of the mean and the variance of $n_r$ are given in Eqs (\ref{meanf},\ref{varfast}) and Eqs (\ref{means},\ref{varslow}) }
\end{table*}

We are now able to express the state of the system at time $t+\tau$, $e(t+\tau)$, as a function of the state of the system at time $t$ for a fast fluctuating extrinsic noise ($\tau\gg\tau_s$) and a slow fluctuating extrinsic noise ($\tau\ll\tau_s$),
\begin{equation}
\label{prelangevin}
e(t+\tau)=e(t)+{\bf M}e(t)\tau+ {\bf K}\tau+ {\bf \Sigma}(t)\sqrt{\tau}
\end{equation}
where the conservation law has been used and where the matrices ${\bf M}$, ${\bf K}$ and the vector ${\bf \Sigma}(t)$ are given by
\begin{eqnarray}
{\bf M}&=&\left( \begin{array}{cc} -d_r-2k & 0 \\ d_r-k & -3k \end{array}\right), \quad {\bf K}=\left(\begin{array}{c}kN \\kN \end{array}\right), \\ {\bf \Sigma}(t)&=&\left(\begin{array}{c} \Sigma_1(t) \\ \Sigma_2(t) \end{array}\right)
\end{eqnarray}
with
 \begin{eqnarray}
\Sigma_{1}(t)&=&-\sqrt{v_{r}^2}\delta^{E \rightarrow ES}+\sqrt{kN_{ES}}\delta^{ES \rightarrow E}\nonumber \\
&-&\sqrt{kN_{E}}\delta^{E \rightarrow EP}+\sqrt{kN_{EP}}\delta^{EP \rightarrow E},\\
\Sigma_{2}(t)&=&\sqrt{v_{r}^2}\delta^{E \rightarrow ES}-\sqrt{kN_{ES}}\delta^{ES \rightarrow E}\nonumber \\
&-&\sqrt{kN_{ES}}\delta^{ES \rightarrow EP}+\sqrt{kN_{EP}}\delta^{EP \rightarrow ES}.
\end{eqnarray}
The expressions of $d_r$ and $v_r$ are given in Table 1. The dependence in $t$ of $N_i(t)$ is omitted for the ease of notation.
The random variables $\delta^j$ can be rewritten as $\delta^j(t)$; $\delta^i(t)$ and $\delta^j(t')$ will be independent either if $i \neq j$, $t \neq t'$.  We use the property that a stochastic variable of vanishing mean and variance $1/\sqrt{\tau}$  is a white noise when $\tau\rightarrow 0$, we replace the notation $\delta^i(t)/\sqrt{\tau}$ by $\Gamma^i(t) $ and $\tau$ by $dt$ and we rewrite Eq (\ref{prelangevin}) for the fast and the slow fluctuating noise. In both cases, the multiplicative noises have to be understood according to the Ito convention \cite{gardiner}.

We find for a NESS perturbed by a fast relaxing extrinsic noise,
\begin{equation}
\label{activextrinlangevin}
\frac{d e(t) }{dt} = {\bf M}_fe(t)+ {\bf K} + {\bf \Gamma}(t),
\end{equation}
where
\begin{eqnarray}
\label{Mfgamma}
{\bf M}_f=\left(\begin{array}{cc} -k(2+s_m) &  0 \\ k(s_m-1) & -3k \end{array}\right)
  \quad
 {\bf \Gamma}(t)&=&\left(\begin{array}{c}
\Gamma_{E}(t) \\ \Gamma_{ES}(t) \end{array}\right), 
\end{eqnarray}
with 
\begin{eqnarray}
\label{gammaE}
\Gamma_{E}(t)&=&-\sqrt{v_r^2}\Gamma^{E \rightarrow ES}(t)+\sqrt{kN_{ES}}\Gamma^{ES \rightarrow E}(t)\nonumber \\
&-&\sqrt{kN_{E}}\Gamma^{E \rightarrow EP}(t)+\sqrt{kN_{EP}}\Gamma^{EP \rightarrow E}(t)\\
\label{gammaES}
\Gamma_{ES}(t)&=&\sqrt{v_r^2}\Gamma^{E \rightarrow ES}(t)-\sqrt{kN_{ES}}\Gamma^{ES \rightarrow E}(t)\nonumber \\
&-&\sqrt{kN_{ES}}\Gamma^{ES \rightarrow EP}(t)+\sqrt{kN_{EP}}\Gamma^{EP \rightarrow ES}(t)
\end{eqnarray}
in which the time dependence of the particle number has been omitted to simplify the expressions.
Equation (\ref{activextrinlangevin}) is a Chemical Langevin Equation  for a chemical system driven in a non-equilibrium state by a noisy driving force. Its deterministic part is identical to the one describing an unperturbed NESS but the amplitude of the white noise is increased. The extrinsic noise plays the role of an additional thermal force. 
The particular case of a system maintained in a NESS by a constant driving force is obtained by 
writing $v_r^2=kN_E(t)s_m$ in Eqs. (\ref{gammaE},\ref{gammaES}).

We now consider the stochastic equation obtained for a slow extrinsic noise $(\tau_s\gg \tau)$. In this case, we get
\begin{equation}
\label{activextrinlangevinslow}
\frac{d e(t) }{dt} = {\bf M}_se(t)+ {\bf K} + {\bf \Gamma}(t),
\end{equation}
where
\begin{eqnarray}
\label{Ms}
{\bf M}_s=\left(\begin{array}{cc} -k(2+s_m+\delta s(t)) &  0 \\ k(s_m+\delta s(t)-1) & -3k \end{array}\right).
\end{eqnarray}
The expression of ${\bf \Gamma}(t)$ given in Eqs (\ref{Mfgamma}-\ref{gammaES}) and Table 1. This equation is not a Chemical Langevin Equation because of the stochastic rates appearing in the matrix {\bf M}$_s$. 
The fluctuating part of $s(t)$ generates an additional multiplicative force.
 
 We obtain for the constant driving force, the fast relaxing fluctuations and the slow relaxing fluctuations, the following correlation functions
\begin{eqnarray}
\label{correl0}
\langle \delta N_E(t) \delta N_E(t') \rangle_c&=&\frac{s_m+1}{s_m+2}\langle N_E\rangle e^{-(t-t')/\tau_r},\\
\label{correlf}
\langle \delta N_E(t) \delta N_E(t')\rangle_f&=&\left(1+k\frac{\langle N_E \rangle}{s_m+1} \sigma_s\tau_s \right) \nonumber \\ &\times& \langle \delta N_E(t) \delta N_E(t') \rangle_c,\\
\label{correls}
\langle \delta N_E(t) \delta N_E(t')\rangle_s&=&\langle \delta N_E(t) \delta N_E(t') \rangle_c\nonumber\\
&+&\sigma_s^2\frac{k^2 \tau_r^2\tau_s}{(\tau_s^2-\tau_r^2)}\langle N_E\rangle ^2 \nonumber \\
& &\left(\tau_se^{-(t-t')/\tau_s}-\tau_r e^{-(t-t')/\tau_r}\right),
\end{eqnarray}
where $t>t'$, $\tau_r=1/k(2+s_m)$ is the relaxation time of $N_E(t)$ for a non-perturbed NESS and $\delta N_E(t)$ is the difference between $N_E(t)$ and its stationary value  that is given in Eq. (\ref{station}) for the fast relaxing driving force and in Eq. (\ref{NEstatslow}) for the slow process. 
The derivation of an approximate solution of Eq. (\ref{activextrinlangevin}) is performed in Appendix B.  Eq. (\ref{activextrinlangevinslow})  is solved in Appendix C
 in the particular case of weak fluctuation amplitudes ($\Delta \ll 1$).  
The correlation functions are represented in Fig 2. The expressions given in Eqs (\ref{correlf},\ref{correls}) are compared to 
numerical results and the agreement is good.

Lets consider first the correlation function given in Eq. (\ref{correl0}) obtained in the absence of extrinsic noise. The system is at equilibrium for $s_m=1$. Otherwise, the chemical cycle is in a NESS characterized by the value of $s_m$ and is associated with a reactive current.  The variance $\langle \delta N_E(0) \delta N_E(0) \rangle_c$ and the characteristic time  $\tau_r=1/k(2+s_m)$ of the correlation function of $N_E(t)$ are functions of $s_m$. 

When a given NESS ($s_m$) is perturbed by a fast fluctuating noise, ($\tau_s \ll \tau$), the correlation function given in Eq. (\ref{correlf}) remains proportional to one obtained in the absence of noise, the characteristic time is unchanged and only the variance is increased by the extrinsic noise.

A slow relaxing noise perturbing a NESS generates a stochastic multiplicative force, as shown in Eq. (\ref{activextrinlangevinslow}), whose dynamics interferes with the  relaxation of the chemical system. The stationary state of the system is modified (see Appendix C). The correlation function of $N_E$ is the sum of the correlation of the unperturbed NESS and of a contribution associated with the characteristic time of the extrinsic noise $\tau_s$ and of the chemical reaction $\tau_r$. The amplitude of the corrective term scales in $N^2$ and depends on the properties of the extrinsic noise and can be negligible or dominant when compared to the unperturbed NESS correlation function depending on the parameters characterizing $s(t)$.

\begin{figure}
\includegraphics[scale=0.3]{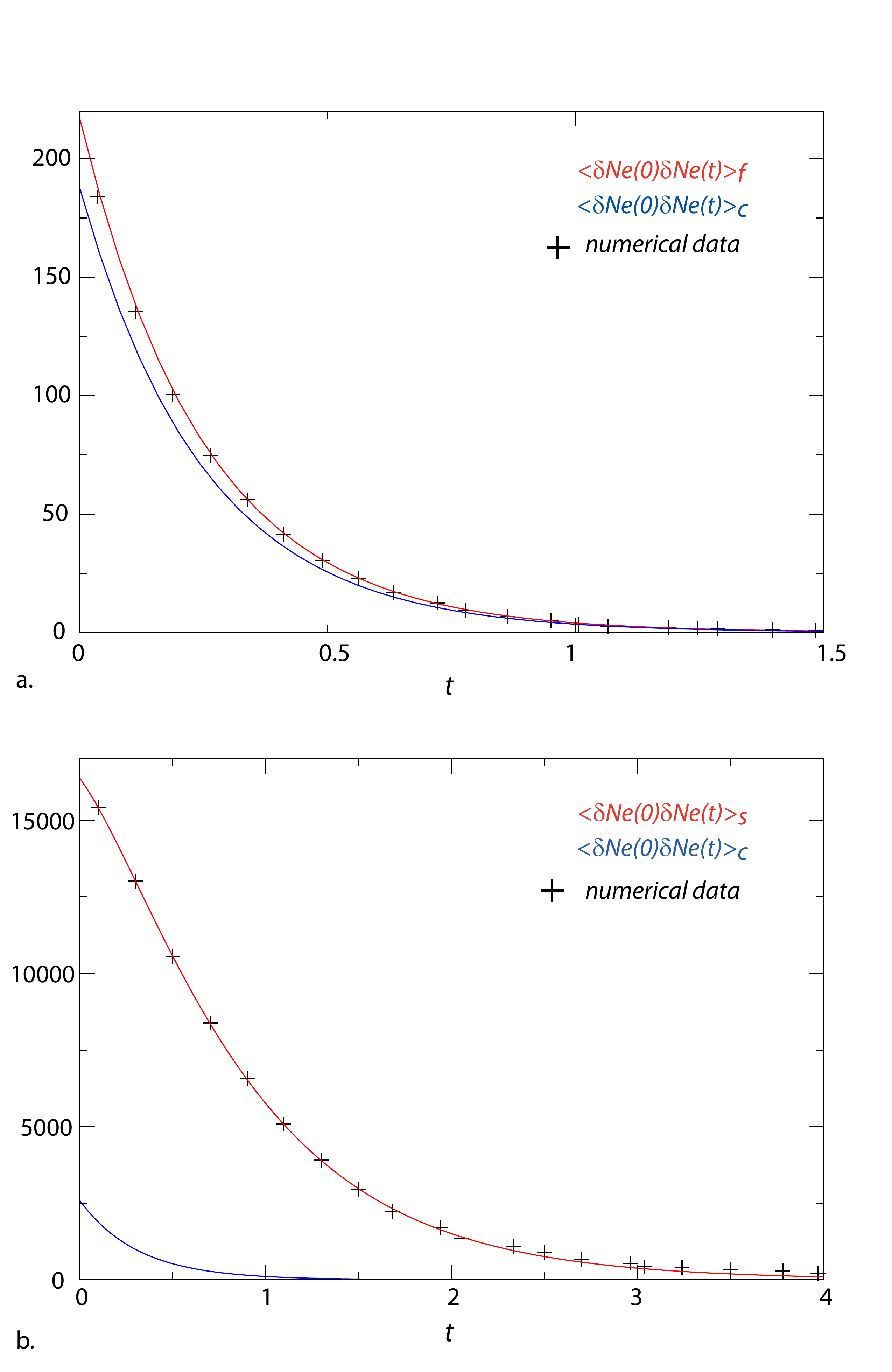}
\caption{Correlation function in the presence of extrinsic noise. {\bf a}. $\langle N_E(t) N_E(0) \rangle$ for a system maintained in a NESS by a constant driving force and a fast fluctuating driving force $( \tau_s \ll \tau )$. The analytic expression given in  (\ref{correlf}) (red line) is compared to numerical results (+) for the set of parameters in dimensionless units ($N$=1000, $s_m=2$ $\tau_r=0.25$, $\tau_s=10^{-4}$, $\sigma_s=0.2$). The correlation function without extrinsic noise given in Eq. (\ref{correl0}) is represented in blue.  {\bf b}. $\langle N_E(t) N_E(0) \rangle$ for a system maintained in a NESS by a constant driving force and by a fast fluctuating driving force $( \tau_s \ll \tau )$. The analytic expression given in  (\ref{correls}) (red line) is compared to numerical results (+) for the set of parameters ($N$=12000, $s_m=1.2$ $\tau_r=0.3125$, $\tau_s=0.72$, $\sigma_s=0.06$). The correlation function without extrinsic noise Eq. (\ref{correl0}) is represented in blue.}
\end{figure}

\section{Modified Fluctuation-Dissipation Theorem in the presence of extrinsic noise}

The response to a perturbation or the fluctuations of a system in an equilibrium state contain the same information and one can equally collect one or the others depending on experimental constraints. The analysis of FCS data is based on this equivalence and the seminal paper developing the theoretical basis of this method treats the fluctuations as macroscopic perturbations that relax according to chemical deterministic equations \cite{elson1974}. 
The fluctuation-dissipation theorem links the macroscopic response to a perturbation and the correlation functions of the unperturbed state. A modified version of the theorem establishes a similar relation for systems in a non-equilibrium state \cite{seifert2012}.

The results of the previous section show that the correlation functions of a system in a NESS perturbed by a stochastic force depend on the dynamic properties of the extrinsic noises. In certain cases, the correlations inform us on the environmental noise and not anymore on the system dynamics. The breakdown of the modified fluctuation-dissipation theorem is a good criterion to distinguish these two regimes. In this section we first derive the fluctuation-dissipation theorem and its modified version for a system in a NESS in the absence of stochastic driving ($\Delta=0$). Then, we study how this relation is violated in the presence of extrinsic noise.    

We consider a chemical network in a equilibrium state $(\lambda_{eq}=0,\Delta_{eq}=0)$ that is perturbed at time $t'$  by a modification
$\lambda(t)=\lambda_{eq}+h(t)$ of the normalized chemical potential of the species S.  
The response function of $N_E(t)$ associated with this variation is defined as
\begin{eqnarray}
\chi_{N_E,\lambda}(t,t')&=&\left.\frac{\delta\langle N_E(t) \rangle}{\delta h (t') }\right|_{h(t)=0},
\end{eqnarray}
with $t >t'$. Solving the deterministic part of Eq (\ref{activextrinlangevin})  in the frame of the linear response theory, one obtains
\begin{equation}
\chi_{N_E,\lambda}(t,t')=-k\langle N_E\rangle e^{-3k(t-t')}.
\end{equation}
The fluctuation-dissipation theorem states that the response function of $N_E(t)$  can be expressed as the correlation of this observable with the conjugated variable of the perturbation $\lambda$ with respect to energy, {\it i. e.}  $s(t)$. This gives
\begin{eqnarray}
\label{chiemu}
  \chi_{N_E,\lambda}|_{\lambda=0}(t,t')&=&\frac{d}{dt'}\langle N_E(t)s(t')\rangle \\
\nonumber.
\end{eqnarray}
where the reaction flux $d s(t')/dt'$ is equal to $-kN_E(t')+kN_{ES}(t')$. 

For a system maintained out-of-equilibrium by stationary constraints, in our case $(\lambda\neq 0, \Delta=0)$, generalizations of the fluctuation-dissipation theorem have already been obtained in a large number of cases \cite{gomezsolano2009,prost2009,verley2011}.   
The first modified fluctuation-dissipation relation for driven Markovian dynamics on a network of chemical states was proposed by U. Seifert \cite{seifert20102}. The response function of a state concentration was expressed as its correlation with appropriate currents.  A short time later, this relation was rederived for a molecular motor cycle maintained out-of-equilibrium by a steady concentration of ATP and submitted to mechanical or chemical perturbation \cite{verley2011}. 
We express the currents as functions of $N_E(t)$ and $N_{ES}(t)$ using the expressions derived in refs \cite{seifert20102,verley2011},
\begin{eqnarray}
j(t)&=&-\frac{P_{ES}}{P_E}kN_E(t)+\frac{P_E}{P_{ES}}ks_mN_{ES}(t)\\
\nu(t)&=&\left(ks_m-k\frac{P_{ES}}{P_E}\right)N_E(t),
\end{eqnarray}
$P_i$ being the stationary probability of configuration $i$ given in Eq (\ref{proba}).
The current $j$ can be interpreted as an instantaneous rate of consumption of $s$ and $\nu$ as a local rate of consumption of $s$ \cite{verley2011}. 
The response function for a system maintained in a NESS by a driving force associated to a parameter $\lambda$ is equal to 
\begin{equation}
\chi_{N_E,\lambda}(t,t')=-ks_m \langle N_E\rangle e^{-(2+s_m)(t-t')}.
\end{equation}
The modified fluctuation-dissipation theorem is expressed by
\begin{eqnarray}
\label{MFDT}
\chi_{N_E,\lambda}(t,t')=\langle N_E(t)(j(t')-\nu(t'))\rangle. 
\end{eqnarray}
As expected, the correlation functions calculated by solving the Chemical Langevin Equation obtained in the absence of extrinsic noise satisfy Eq. (\ref{MFDT}). 

This relation is not valid anymore for a NESS perturbed by an extrinsic noise. To quantify the extent to which the modified fluctuation-dissipation theorem is violated by an additional external noise, we introduce an effective temperature 
\begin{eqnarray}
\label{Tefdef}
T_{eff}(t)&=&\frac{\langle N_E(t)(j(0)-\nu(0))\rangle}{k_b\chi_{N_E,\Delta\mu_s}}
\end{eqnarray}
with $\lambda=\Delta\mu_s/k_bT_0$ and $T_0$ is the temperature of the thermostat.
In the absence of extrinsic noise, $T_{eff}$ is constant and equal to $T_0$. For a NESS perturbed by a fast relaxing noise, correlations and response keep the same time dependence. Using the expressions of the correlation functions given in Eq. (\ref{correlf}) and in Appendix B and injecting it in Eq (\ref{Tefdef}), one sees that in this case, the effective temperature will be constant and equal to 
\begin{eqnarray}
\label{Teff}
T_{eff,f}&=&T_0\left(1+2k\frac{\langle N_E \rangle}{2s_m+1}\sigma_s^2\tau_s \right).
\end{eqnarray}
 The way a noise keeps a NESS away for the temperature $T_0$ depends on the unperturbed NESS ($\langle N_E \rangle$, $s_m$) and on the noise parameters ($\tau_s, \sigma_s)$. 
A NESS perturbed by a noise that fluctuates slower than $\tau$ will be associated with a time dependent temperature. However, if the characteristic time of the noise remains smaller than the one of the chemical cycle $(\tau_s < \tau_r)$, $T_{eff}(t)$ will tend to a constant value for a time $t > \tau_s$. Its value is obtained by replacing in Eq. (\ref{Tefdef})  the correlation functions by their the expressions given in Eq. (\ref{correls}) and in Appendix C. One gets 
\begin{equation}
\label{Tefs}
T_{eff,s}=T_0\left(1+2\frac{2+s_m}{2s_m+1}\frac{\sigma_s^2k^2\tau_r^3\tau_s}{\tau_r^2-\tau_s^2}\langle N_E\rangle\right)
\end{equation}
For $\tau_s > \tau_r$, $T_{eff}(t)$ diverges for large $t$, the response vanishing faster than the correlations. These cases are illustrated in Figure 3. We consider a chemical system that is perturbed either by a fast extrinsic noise ($\tau_s \ll \tau$) or a slow extrinsic noise ($\tau_s\gg \tau$). We determine numerically the effective temperature given in Eq. (\ref{Tefdef}) by solving the master equation associated to the two systems and plot the ratio of this function and of the corresponding constant effective temperature. As expected, the ratio is constant and equal to one in the first situation and converges to 1 for $t > \tau_s$ in the second situation.
\begin{figure}
\includegraphics[scale=0.3]{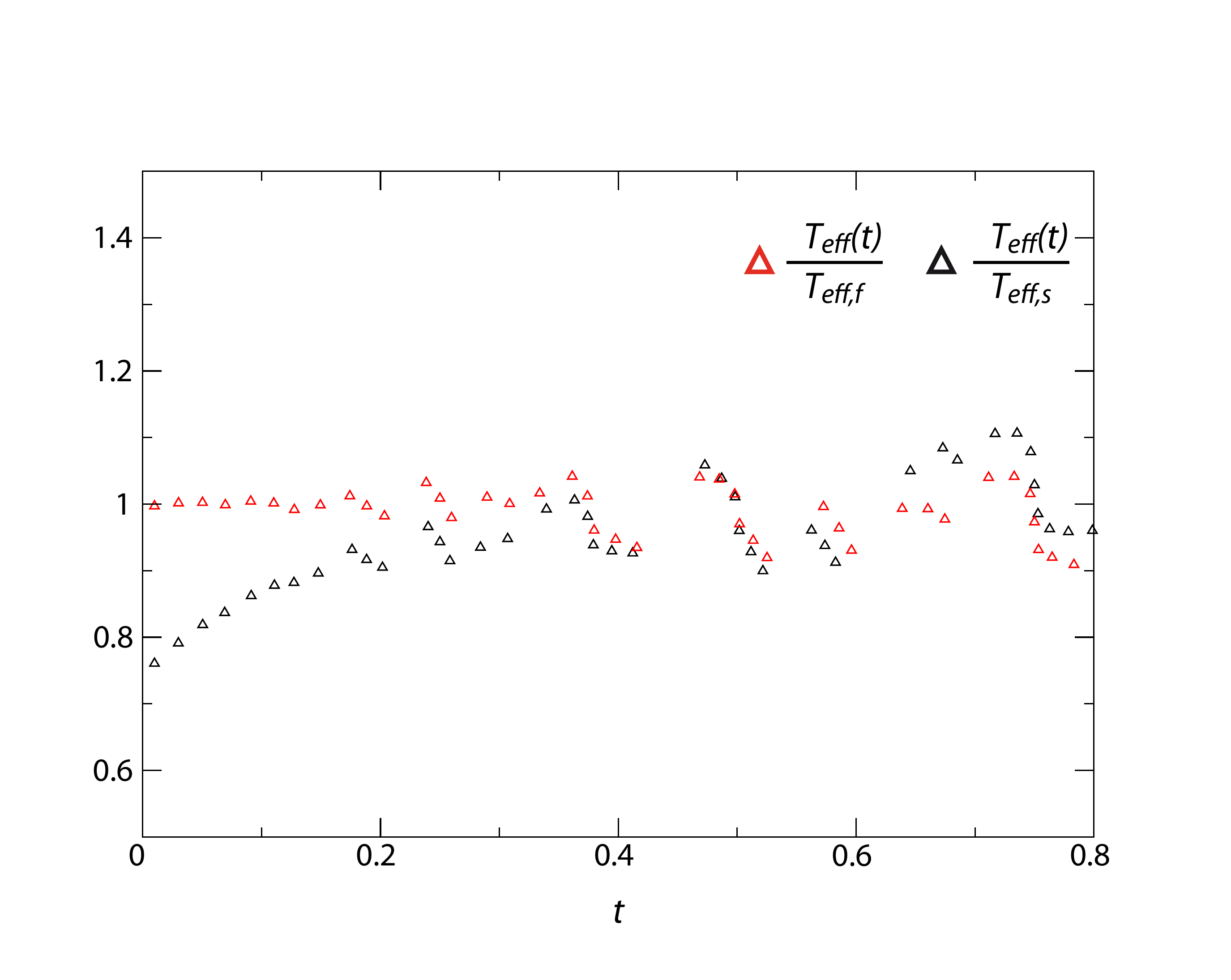}
\caption{Effective temperature in the presence of an extrinsic noise.
We plot the ratio $\frac{Teff(t)}{T_{eff,f}}$ and $\frac{T_{eff}(t)}{T_{eff,s}}$ as a function of $t$ for the chemical system characterized by the parameter set $N=12000, \tau_r=0.3125, s_m=1.2$ and submitted to the noise $\tau_s=0.01,\sigma_s=0.12$ (\textcolor{red}{$\bigtriangleup$}) and $\tau_s=0.1,\sigma_s=0.12$ ($\bigtriangleup$).  The expressions of $T_{eff}(t)$,  $T_{eff,f}$ and $T_{eff,s}$ are given in Eqs. (\ref{Tefdef}-\ref{Tefs}).}
\end{figure}

 An effective 
temperature has been already experimentally measured for a biochemical reaction, the folding of a single short-DNA hairpin driven by a fluctuating force \cite{dieterich2015}. In agreement with our results, the experimentalists observed that correlations and response are proportional when the fluctuating part of the driving force is fast enough. 

To conclude this section, one can see that the fluctuations of a system in a NESS perturbed by an extrinsic noise are not necessary related to its response function. The link between the correlation functions and response breaks in a different manner depending on the dynamic properties of the system and of the noise. This can be illustrated by the introduction of an effective temperature. For the less favorable case, the correlation function time-dependence is controlled by the extrinsic noise.


\section{Conclusion and discussion}
 The experimental evidence that gene expression variability is not only determined by intrinsic fluctuations but also depends on environmental noise was provided 15 years ago\cite{elowitz2002}. It is now clear that extrinsic noises may have an important effect on the statistics of cellular events. Important theoretical works studied their influence on the dynamics of
network motifs that compose gene regulation pathways \cite{alon2007}. First, extrinsic noises were supposed to perturb the input signal received by the chemical networks. 
More recently, the effect of fluctuating chemical rates on the properties of switch circuits were considered \cite{assaf2013,roberts2015} .
The dynamics of such systems is described  by master equations or Fokker-Planck equations that can be approached by Hamiltonian systems. Statistical properties of generic chemical networks in which a rate is perturbed by a bounded extrinsic noise were numerically studied \cite{caravagna2013}.  

In this paper, we considered the fluctuations of reactive species concentrations in conditions that model {\it in vivo} environment. To do so, we studied a chemical cycle in a NESS perturbed by a stochastic force. The system can be far from equilibrium and subject to extrinsic noises. We have adapted the elegant approach proposed by Gillespie to derive stochastic equations describing the dynamics of the network. This consists in coarse-graining the dynamics of the system on a small macroscopic time scale $\tau$ by expressing the number of reactions occurring in each reactive channel by normal variables with identified mean and variance.  This is a very convenient framework to investigate the effect of fluctuating rates. A fast fluctuating rate mimics an extra thermal force, whereas a slow one generates an extra multiplicative force. To quantify the corrections induced by the extrinsic noise on the correlation functions we have studied the breakdown of the modified fluctuation-dissipation theorem and introduced an effective temperature   
as the ratio of correlations and response. This ratio is constant when the extrinsic noise does not modifies the relaxation time of the fluctuations, {\it i. e.} when the noise relaxes faster as the chemical system.

Spectroscopy methods such as Fluorescence Correlation Spectroscopy (FCS) register the {\it in vivo} fluctuations of concentrations and give access to correlation functions for targeted species. This observable is used to discriminate between different physical processes that can govern the fluctuations of the target (passive or active diffusion, reaction, ...) and to evaluate the associated parameters.   In this work, we show that an extrinsic noise inducing a fluctuating rate can drastically perturb the correlation functions of the chemical system.  Many ingredients such as diffusion or migration are missing in this model, however we believe that this work is a step toward the derivation of a theoretical framework well adapted to {\it in vivo} conditions.

\section{Acknowledgment}
We would like to thank R. Chetrite and J. Ranft for discussions and revisions.

\appendix
\renewcommand{\thefigure}{A\arabic{figure}}

\setcounter{figure}{0}
\renewcommand{\thetable}{A\arabic{table}}

\setcounter{table}{0}
\section{Probability distribution of reaction number ${\rm E}\rightarrow {\rm ES}$ during $\tau$}
We evaluate the probability distribution of $n_r^{E\rightarrow ES}$, the number of reactions
${\rm E} \rightarrow {\rm ES}$, that occur during the time $\tau$, when the substrate concentration follows an Ornstein-Uhlenbeck process whose dynamics is governed by Eq. (\ref{O-U}). The time $\tau$ is a mesoscopic timescale chosen so that the conditions given in Eqs. (\ref{cond1},\ref{cond2}) are satisfied for the five channels associated with constant rates 

We first consider an Ornstein-Uhlenbeck process that relaxes fast compared to this timescale, {\it i. e.} $\tau_s \ll \tau$. In this case, the stochastic variables $s(t)$ and  $k_a(t)$ defined in Eqs. (\ref{O-U},\ref{ka}) fully explore their stationary distribution on $\tau$. Providing that the condition given in Eq. (\ref{cond1}) is satisfied also for the channel E$\rightarrow$ ES, the number of reactions $n_r^{E\rightarrow ES}$ is a Poisson random variable of rate $\Lambda^{E\rightarrow ES}=kN_E(t)\int_0^{\tau}s(t) dt $ which is itself a stochastic variable. 
The fluctuating contribution $\int_0^{\tau}s(t)dt$ is the integral of a Ornstein-Uhlenbeck variable and is a normal stochastic variable \cite{gillespie1996} with mean value $ s_m\tau $ and variance $2\sigma_s^2\tau_s\tau$.
The rate $\Lambda^{E\rightarrow ES}$ is thus characterized by the following probability law 
\begin{equation}
\label{Plambda}
\mathcal{P}(\Lambda^{E\rightarrow ES})=\frac{1}{\sqrt{2\pi}\sigma_\Lambda}e^{-\frac{(\Lambda^{E\rightarrow ES}-\langle \Lambda \rangle)^2}{2\sigma_s^2}},
\end{equation}
with
\begin{equation}
\langle \Lambda \rangle =kN_E(t)\langle \int_0^{\tau}s(t) dt \rangle=kN_E(t) s_m \tau
\end{equation}
and
\begin{eqnarray}
\sigma_{\Lambda}^2&=&\langle (kN_E(t)\int_0^{\tau}s(t)dt-kN_E(t)\langle s\rangle\tau)^2\rangle \nonumber \\&=&2k^2N_E(t)^2\sigma^2_s\tau_s\tau
\end{eqnarray}
in the limit $\tau\gg \tau_S$.
As a consequence, the number of reactions $n_r^{E \rightarrow ES}$ follows a probability law that can be written as
\begin{equation}
\label{convol}
P(n_r)=\int_{-\infty}^{\infty}d \Lambda^{E\rightarrow ES} \mathcal{P}(\Lambda^{E\rightarrow ES}) P_{\Lambda}(n_r),
\end{equation} 
where $P_{\Lambda}(n_r)$ is the Poisson distribution associated with the rate $\Lambda$.
All the moments of $P(n_r)$  can easily be calculated, as they all take the form of Gaussian integrals. They can be approximated by the moments of the normal distribution 
\begin{equation}
\label{probataupetitapprox}
\mathcal{N}_f(n_r)=\frac{1}{\sqrt{2\pi}\sigma_r}e^{-\frac{(n_r-\langle n_r\rangle)^2 }{2\sigma_r^2}},
\end{equation}
with 
\begin{eqnarray}
\label{meanf}
\langle n_{r} \rangle_f=kN_E s_m \tau, \\ 
\label{varfast}
\sigma_{r,f}^2=\langle n_r \rangle + 2k^2 N_E^2\sigma^2_s\tau_s\tau,
\end{eqnarray}
in the limit $\langle n_r \rangle \gg 1$, which is equivalent to the condition given in Eq. (\ref{cond2}). The moments of the distribution given in Eq. (\ref{probataupetitapprox}) are equal to ones of the distribution given in Eq. (\ref{convol}) up to a negligible correction in $1/\langle\Lambda\rangle^2=1/\langle n_r \rangle^2$. The stochastic variable $n_r^{E\rightarrow ES}$ follows the normal distribution given in Eq. (\ref{probataupetitapprox}) if the conditions given in Eqs. (\ref{cond1},\ref{cond2}) are fulfilled.

We are now interested in the probability distribution of $n_r^{E\rightarrow ES}$ 
when the relaxation time $\tau_s$ of the fluctuations of $s(t)$ is much larger than the time $\tau$. 
$s(t)$ and $k_a(t)$ are assumed to be constant on $\tau$. For a fixed value of $s(t)$, we are back to the case of passive channels with constant kinetic rates. The number of reaction is a Poisson random variable characterized by a rate $\Lambda^{E\rightarrow ES}=kN_Es(t)\tau$, providing that $kN_E(t)s(t)\tau \ll N_E(t)$, {\it i. e.} that the condition given in Eq. (\ref{cond1}) is fulfilled. The mean value and the variance of $n_r^{E\rightarrow ES}$  are thus $\langle n_r^{E\rightarrow ES}\rangle=kN_E(t)s(t)$ and $\sigma_r^2=kN_E(t) s(t)\tau$. This Poissonian law can be approximated by a normal distribution if the condition given in Eq. (\ref{cond2}) is satisfied.
We consider an ensemble of $n$ time intervals $\tau$ associated to an ensemble of values of reaction numbers $n_r$ and to a set of values $s_i$, ${i=1, ..., n}$, of  $s(t)$, such that $(s_i)$ reaches its stationary distribution given in Eq. (\ref{PS}). The number of reactions $n_r$ is a stochastic variable following the normal probability distribution
\begin{equation}
\label{probataugrandapprox}
\mathcal{N}_s(n_r)=\frac{1}{\sqrt{2\pi}\sigma_r}e^{-\frac{(n_r-\langle n_r\rangle)^2 }{2\sigma_r^2}},
\end{equation}
with  
\begin{eqnarray}
\label{means}
\langle n_{r}\rangle_s &=& kN_E(t) s_m \tau, \\
\label{varslow}
\sigma_{r,s}^2&=&kN_E(t) s_m \tau+k^2N_E^2\sigma_s^2\tau.
\end{eqnarray} 

In both cases, the reaction number $n_r$ is a stochastic variable associated with a normal distribution which properties are summed in Table A. I. The probability distributions of $n_r^{E\rightarrow ES}$ are also obtained numerically. The numerical results are plotted and compared to analytic results in the Fig. A1. The agreement is very good. 
\begin{table*}
\center
\renewcommand{\arraystretch}{1.6}
\setlength{\tabcolsep}{1cm}
\begin{tabular}{ccc}
\hline
 & $n_{\tau}$ & $\sigma_r^2$  \\
\hline
\hline
$\Delta=0$ & $kN_E(t)s_m \tau$  & $kN_E(t)s_m \tau$  \\
\hline
$\tau_s \ll \tau$ & $kN_E(t)s_m \tau$ & $kN_E(t)s_m \tau$  + $2k^2N_E^2(t)\sigma^2_s\tau_s\tau $  \\
\hline
$\tau_s \gg \tau$ &$kN_E(t)s_m \tau$  & $ kN_E(t)s_m \tau $ + $k^2N_E^2(t)\sigma_s^2\tau^2 $ \\
\hline 
\end{tabular}
\caption{Mean value and variance of the reaction number $n^{E\rightarrow ES}_r$ during a mesoscopic time $\tau$. The values are given for a constant driving force, for a driving force that relax much faster than $\tau$ and a driving force that relaxes much slower than $\tau$.}
\end{table*}
\begin{figure}
\includegraphics[scale=0.3]{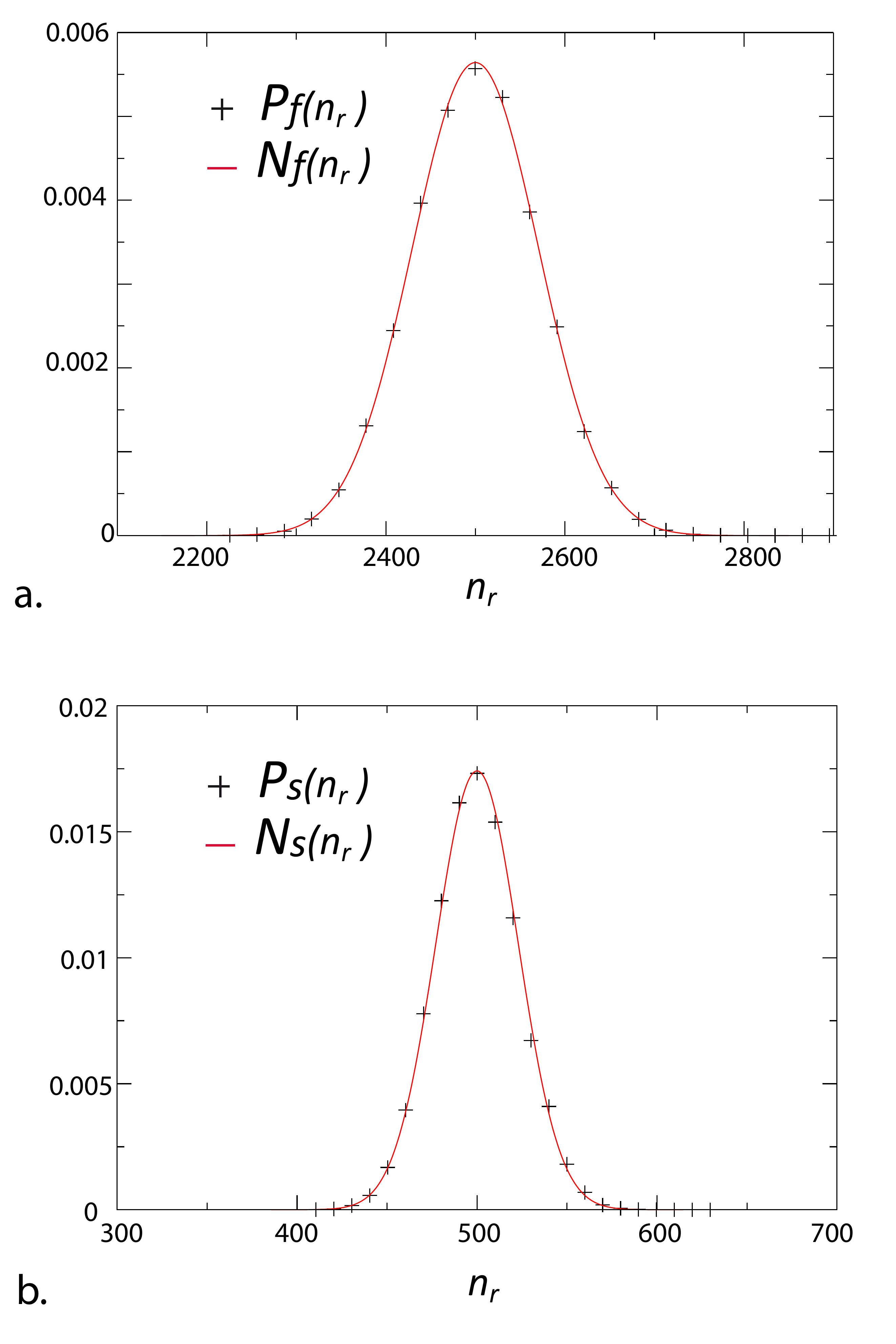} 
\caption{Probability distribution $P(n_r)$ of the reaction number $n_r$ of E $\rightarrow$ ES during a time $\tau$. 
Fig. {\bf a}, Fast Ornstein-Uhlenbeck process. The figure represents the probability distribution for the number of reactions of E$\rightarrow$ ES during the time $\tau$ obtained numerically ($+$) and theoretically (red line). The analytic expression is given in Eq. (\ref{probataupetitapprox}). Both theoretical and numerical curves are obtained for $N=1000000$, $k=1$, $\tau=0.005$, $s_m= 2$, $\sigma=800$, $\tau_s=0.0001$ in dimensionless units.
Fig. {\bf b}, Slow Ornstein-Uhlenbeck process. The figure represents the probability distribution for the number of reactions of E$\rightarrow$ ES during the time $\tau$ obtained numerically ($+$) and theoretically (red line). The analytic expression is given in Eq. (\ref{probataugrandapprox}). Both theoretical and numerical curves are obtained for $N=1000000$, $k=1$, $\tau=0.001$, $s_m=2$, $\sigma=0.08$, $\tau_s=0.01$ in dimensionless units.}
\end{figure}

\section{Solution of the Chemical Langevin Equation in the case of a fast relaxation driving force.}
In this part we derive an approximate solution of the Chemical Langevin Equation (\ref{activextrinlangevin}) obtained when $s(t)$ relaxes fast when compared to the incremental time $dt$.
 To do so, we approximate the time dependent diffusion coefficients of Eq. (\ref{activextrinlangevin}) by their mean values and the noise term ${\bf \Gamma}(t)$ by ${\bf \Gamma^s}(t)$ with
\begin{eqnarray}
\label{GammaElin}
\Gamma^s_{E}(t)&=&-\sqrt{\langle v_r^2 \rangle}\Gamma^{E \rightarrow ES}(t)+\sqrt{k\langle N_{ES}\rangle}\Gamma^{ES \rightarrow E}(t)\nonumber \\
&-&\sqrt{k\langle N_{E}\rangle}\Gamma^{E \rightarrow EP}(t)
+\sqrt{k\langle N_{EP}\rangle}\Gamma^{EP \rightarrow E}(t)\\
\label{GammaESlin}
\Gamma^s_{ES}(t)&= &\sqrt{\langle v_r^2\rangle }\Gamma^{E \rightarrow ES}(t)-\sqrt{k\langle N_{ES}\rangle}\Gamma^{ES \rightarrow E}(t)\nonumber \\
&-&\sqrt{k\langle N_{ES} \rangle}\Gamma^{ES \rightarrow EP}(t)\nonumber\\&+&\sqrt{k\langle N_{EP}\rangle}\Gamma^{EP \rightarrow ES}(t).
\end{eqnarray}
with $\langle v_r^2 \rangle = k\langle N_E \rangle s_m+2k^2\langle N_E\rangle^2\sigma_s\tau_s$.
The linearized Chemical Langevin Equation characterizing the dynamics of the cycle given in Fig. 1. is obtained by replacing 
${\bf \Gamma} (t)$ by  ${\bf \Gamma}^s(t)$.
This equation can be written as 
\begin{equation}
\label{ChemLangEqrap}
\frac{d e(t) }{dt} = {\bf M}_fe(t)+ {\bf K} + {\bf \Gamma}^s(t),
\end{equation}
with 
\begin{eqnarray}
\label{Mr}
{\bf M}_f=\left(\begin{array}{cc} -k(2+s_m) & 0 \\ k(s_m-1) & -3k \end{array}\right) 
\end{eqnarray}
and the coefficients of ${\bf \Gamma}^s(t)$ given in Eqs. (\ref{GammaElin},\ref{GammaESlin}).
The Eq. (\ref{ChemLangEqrap}) is solved without difficulties and one gets
\begin{eqnarray}
N_E(t)&=&\langle N_E \rangle+\int_{-\infty}^t dt'\Gamma^s_E(t')e^{k(2+s_m)t'}e^{-k(2+s_m)t},\\
N_{ES}(t)&=&\langle N_{ES} \rangle+\int_{-\infty}^t dt'(-\Gamma^s_E(t')+\Gamma^s_{ES}(t'))e^{3kt'}e^{-3
kt} \nonumber\\
&+&\int_{-\infty}^t dt'\Gamma_E^s(t')e^{k(2+s_m)t'}e^{-k(2+s_m)t},
\end{eqnarray}
with the stationary values
\begin{eqnarray}
\label{NEstat}
\langle N_E \rangle&=&\frac{N}{2+s_m}, \nonumber\\
\label{NESstat}
\langle N_{ES} \rangle&=&\frac{1+2s_m}{3(2+s_m)}N.
\end{eqnarray}
 The corresponding correlation functions are 
\begin{eqnarray}
\label{Nenef}
\langle \delta N_E(t) \delta N_E(t')\rangle&=&\frac{D_{E,E}}{2k(2+s_m)}e^{-k(2+s_m)(t-t')}\\
\langle \delta N_E(t) \delta N_{ES}(t') \rangle &=&\Bigg(\frac{D_{E,E}+D_{E,ES}}{k(5+s_m)}-\frac{D_{E,E}}{2k(2+s_m)}\Bigg)\nonumber \\ & & e^{-k(2+s_m)(t-t')}
\end{eqnarray}
\begin{eqnarray}
\langle \delta N_{ES}(t) \delta N_E(t')\rangle &=& \frac{D_{E,E}+D_{E,ES}}{k(5+s_m)}e^{-3k(t-t')}\nonumber \\&-&\frac{D_{E,E}}{2k(2+s_m)}e^{-k(2+s_m)(t-t')}
\end{eqnarray}
\begin{eqnarray}
\langle \delta N_{ES}(t) \delta N_{ES}(t')\rangle &=&\Bigg( \frac{D_{E,E}+2D_{E,ES}+D_{ES,ES} }{6k}\nonumber \\&-&\frac{D_{E,E} + D_{E,ES}}{k(5+s_m)}\Bigg)e^{-3k(t-t')}\nonumber\\
&+&\Bigg(\frac{D_{E,E}}{2k(2+s_m)}+\frac{D_{E,E}+ D_{E,ES}}{k(5+s_m)}\Bigg)\nonumber \\ & &e^{-k(2+s_m)(t-t')}
\end{eqnarray}
with
\begin{eqnarray}
\label{GEE}
D_{E,E}&=&2k\frac{s_m+1}{2+s_m}N + 2k^2(N_E^s)^2\sigma_s^2 \tau_s \\
\label{GEES}
D_{E,ES} &=&-k\frac{5s_m+1}{3(2+s_m)}N - 2 k^2(N_E^s)^2 \sigma_s^2 \tau_s \\
\label{GESES}
D_{ES,ES}&=& 4k\frac{2s_m+1}{3(2+s_m)}N+2 k^2(N_E^s)^2\sigma_s^2 \tau_s^2
\end{eqnarray}
The correlation functions for a system driven by a constant driving force are obtained by setting $\tau_s$ to zero in Eqs. (\ref{GEE}-\ref{GESES}).

\section{Solution of the stochastic equation obtained in the case of a slow Ornstein-Uhlenbeck process.}
The linearized Chemical Langevin Equation modeling the dynamics of a system driven by an Ornstein-Uhlenbeck process that relaxes slowly when compared to $dt$ can be written as 
\begin{equation}
\label{ChemLangEq}
\frac{d e(t) }{dt} = {\bf M}_se(t)+ {\bf K} + {\bf \Gamma}^s(t),
\end{equation}
with 
\begin{eqnarray}
{\bf M}_s=\left(\begin{array}{cc} -k(2+s_m+\delta s(t)) & 0 \\ k(s_m+\delta s(t)-1) & -3k \end{array}\right), 
\end{eqnarray}
and the coefficients of ${\bf \Gamma}^s(t)$ given in Eqs. (\ref{GammaElin},\ref{GammaESlin}) and Table 1.
The coefficient $\delta s(t)$ in ${\bf M}_s$ generates fluctuations in $N_E(t)$ that scale as $\sigma_sN$, ${\bf \Gamma}^s(t)$ generates fluctuations in $(s_mN)^{1/2}$. We solve Eq. (\ref{ChemLangEq}) for 
 $\sigma_s/s_m \ll 1$, {\it i. e.} for a small amplitude of extrinsic noise. We  develop $e(t)=e^0(t)+\sigma_s/s_m e^1(t)+(\sigma_s/s_m)^2 e^2(t)$ to the second order in $\sigma_s/s_m$ and solve the corresponding equations.

 The expression of Eq. (\ref{ChemLangEq}) at the zeroth order in $\sigma_s/s_m$ gives
\begin{equation}
\label{ChemLangEqslow}
\frac{d e(t) }{dt} = {\bf M^0}e^0(t)+ {\bf K}+{\bf \Gamma}^{s0}(t), 
\end{equation}
with ${\bf M}^0={\bf M}_f$ given in Eq. (\ref{Mr}) and ${\bf \Gamma}^{s0}(t)$ equal to
\begin{eqnarray}
\label{GammaElin}
\Gamma^s_{E}(t)&=&-\sqrt{k\langle N_E \rangle}\Gamma^{E \rightarrow ES}(t)+\sqrt{k\langle N_{ES}\rangle}\Gamma^{ES \rightarrow E}(t)\nonumber \\
&-&\sqrt{k\langle N_{E}\rangle}\Gamma^{E \rightarrow EP}(t)
+\sqrt{k\langle N_{EP}\rangle}\Gamma^{EP \rightarrow E}(t)\\
\label{GammaESlin}
\Gamma^s_{ES}(t)&= &\sqrt{k\langle N_E\rangle }\Gamma^{E \rightarrow ES}(t)-\sqrt{k\langle N_{ES}\rangle}\Gamma^{ES \rightarrow E}(t)\nonumber \\
&-&\sqrt{k\langle N_{ES} \rangle}\Gamma^{ES \rightarrow EP}(t)\nonumber\\&+&\sqrt{k\langle N_{EP}\rangle}\Gamma^{EP \rightarrow ES}(t).
\end{eqnarray}
One gets the stationary values of $N_E(t)$ and $N_{ES}(t)$
\begin{eqnarray}
N_E^0(t)&=&\langle N_E \rangle+\int_{-\infty}^t dt'\Gamma^s_E(t')e^{k(2+s_m)t'}e^{-k(2+s_m)t},\\
N_{ES}^0(t)&=&\langle N_{ES} \rangle+\int_{-\infty}^t dt'(-\Gamma^s_E(t')+\Gamma^s_{ES}(t'))e^{3kt'}e^{-3
kt} \nonumber\\
&+&\int_{-\infty}^t dt'\Gamma_E(t')e^{k(2+s_m)t'}e^{-k(2+s_m)t},
\end{eqnarray}
with the expressions of $\langle N_E \rangle$ and $\langle N_{ES} \rangle$ given in Eqs (\ref{NESstat},\ref{NESstat}).
At the first order in $\sigma_s/s_m$, we solve the following equation
\begin{equation}
\frac{d e^1(t) }{dt} = {\bf M}_fe^1(t)+{\bf C} \epsilon(t)e^{0}(t),
\end{equation} 
 with ${\bf C}=\left(\begin{array}{cc} -k & 0 \\ k & 0\end{array}\right)$, $\delta s(t)=\sigma_s /s_m\epsilon (t)$ and $e^{0s}=\left( \langle N_E\rangle , \langle N_{ES} \rangle \right)$. 
The first-order term $\sigma_s/s_me^1(t)=\left( N_E^1(t), N_{ES}^1(t)\right)$ is thus equal to
\begin{eqnarray}
\label{NE1}
N_E^1(t)&=&-\int_{-\infty}^t dt'k\epsilon(t') N_E^0(t') e^{k(2+s_m)t'}e^{-k(2+s_m)t} \\
\label{NES1}
N_{ES}^1(t)&=&-\int_{-\infty}^t dt'k\epsilon(t') N_E^0(t') e^{k(2+s_m)t'}e^{-k(2+s_m)t} \nonumber \\
&+&\int_{-\infty}^t dt'2 k\epsilon(t') N_E^0(t') e^{3kt'}e^{-3kt}
\end{eqnarray}
The development of Eq. (\ref{ChemLangEq})
at the second order in $\sigma_s/s_m$ gives
\begin{equation}
\frac{d e^2(t)}{dt}= {\bf M}_fe^2(t)-{\bf C}\epsilon(t)e^1(t).
\end{equation}
The mean value $\langle e^2(t) \rangle$ does not vanish, and we calculate it using the expression of $e^1(t)$ given in Eq. (\ref{NE1}). 
We find for the stationnary value at the second order in $\sigma_s/s_m$,
\begin{eqnarray}
\label{NEstatslow}
N_E^s&=&\langle N_E \rangle \left(1+k^2\tau_r^2\sigma_s^2\frac{\tau_s}{\tau_r+\tau_s}\right),
\end{eqnarray}
when use has been made of $\langle \epsilon (t) \Gamma^{i \rightarrow j}(t)\rangle=0$.
The correlation functions are calculated at the second order in $\sigma_s/s_m$ and one finds 
\begin{eqnarray}
\label{correlsA}
\langle \delta N_E(t) \delta N_E(t')\rangle&=&\frac{\sigma_s^2 k^2\tau_r^2\tau_s}{\tau_s^2-\tau_r^2}\langle N_E\rangle^2\nonumber\\
&\times & \left(\tau_se^{-(t-t')/\tau_s}-\tau_r e^{-(t-t')/\tau_r}\right)\nonumber\\ 
&+& \frac{s_m+1}{s_m+2}\langle N_E\rangle e^{-(t-t')/\tau_r},
\end{eqnarray}
with $\delta N_E(t)=N_E(t)-N_E^s$ and $t>t'$.

\bibliographystyle{unsrt}
 \bibliography{heq}
\end{document}